\documentclass[11pt]{article}
\usepackage{moriond,epsfig}
\usepackage{citesort}

\newenvironment{Simlis}[1][$\bullet$]
{\begin{list}{#1}
 {
  \settowidth{\labelwidth}{#1}
  \setlength{\labelsep}{0.5em}
  \setlength{\leftmargin}{0.5em}
  \setlength{\rightmargin}{0em}
  \setlength{\itemsep}{0ex}
  \setlength{\topsep}{0ex}
 }
}
{\end{list}}

\newcommand{\figCtqNewOld}
{
\begin{figure}
\begin{center}
\includegraphics[height=0.25\textheight]{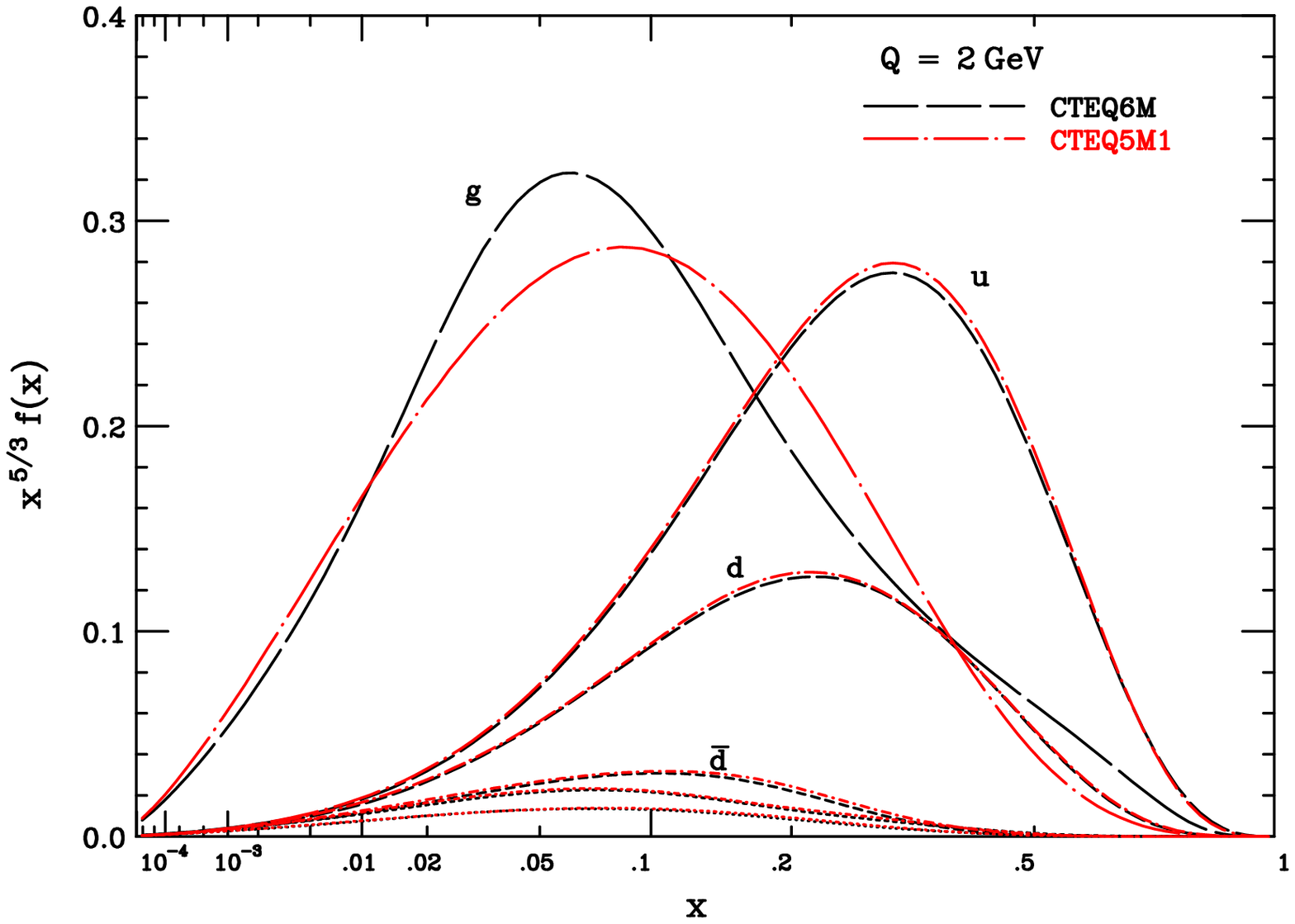}
\end{center}
\caption{Comparison of CTEQ6M to CTEQ5M1 PDFs at $Q=2$ GeV. }
\label{fig:CtqNewOld}
\end{figure}
}
\newcommand{\figLumi}
{
\begin{figure}
\rule{0em}{1ex}\hfill
\includegraphics[height=0.25\textheight]{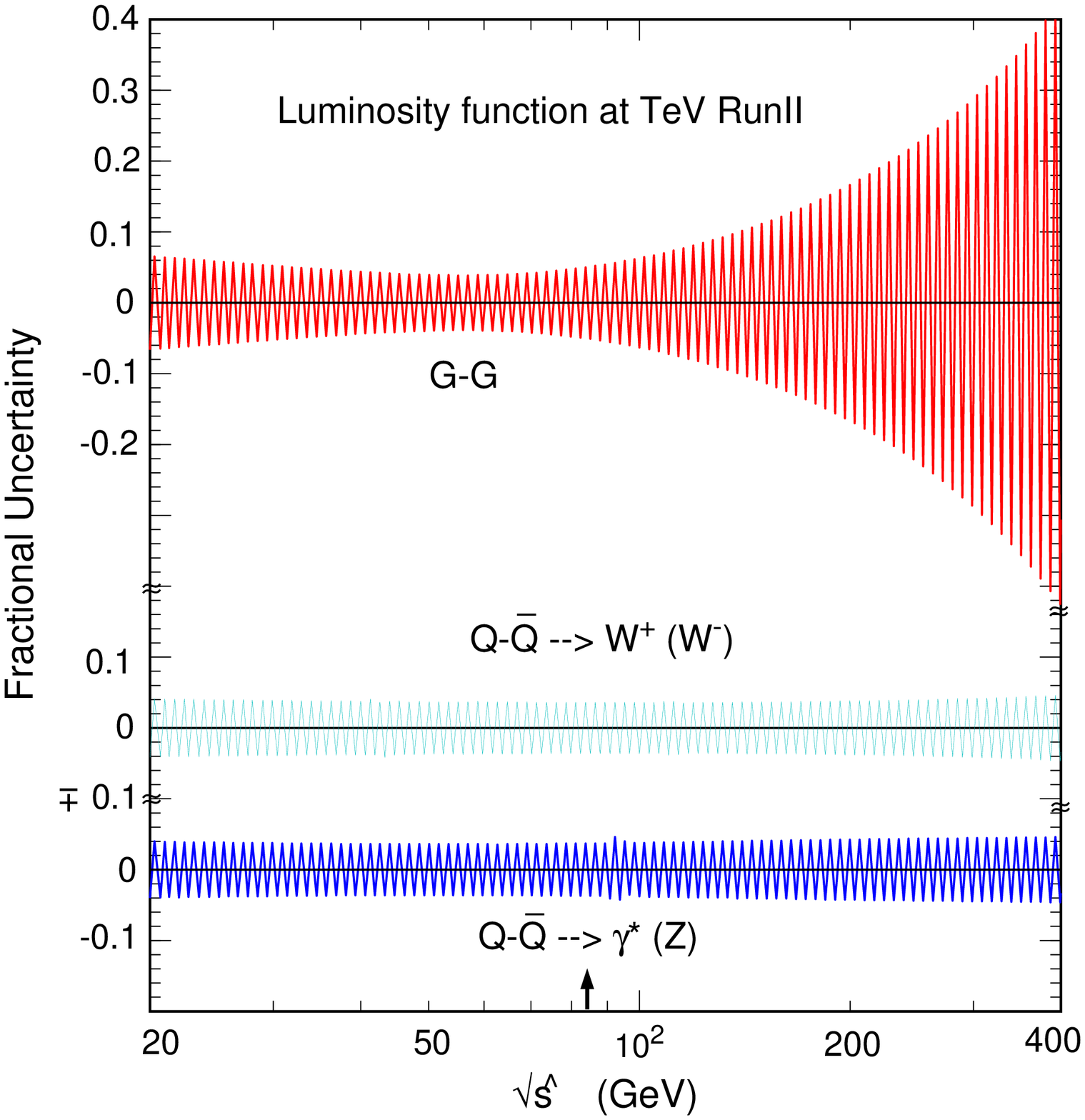}
\hfill\rule{0em}{1ex}\hfill
\includegraphics[height=0.25\textheight]{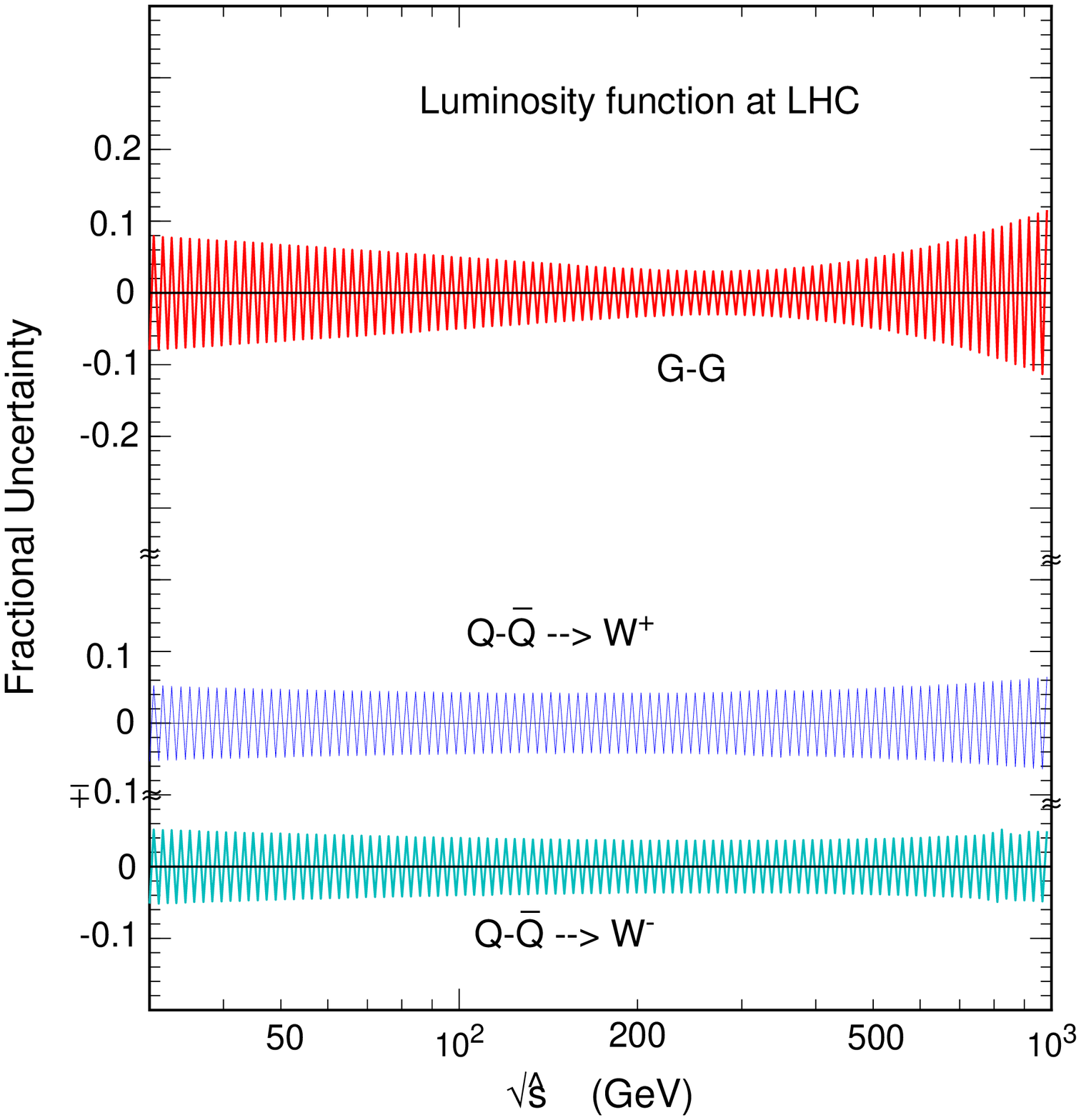}
\hfill\rule{0em}{1ex}

\caption{Uncertainties of the luminosity functions at the
Tevatron and at the LHC.} \label{fig:Lumi}
\end{figure}
}

\begin{document}
\vspace*{4cm}
\title{New Generation of Parton Distributions with Uncertainties \\
from Global QCD Analysis}
\author{Wu-Ki Tung
\footnote{The work reported in this talk has been done in collaboration with
 J.\ Pumplin, D.R.\ Stump, J.\ Huston, H.L.\ Lai, and P. Nadolsky.}}

\address{Department of Physics and Astronomy \\
         Michigan State University \\
         East Lansing, MI 48824 USA }
\maketitle
\abstracts{ A new generation of parton distribution functions with
increased precision and quantitative estimates of uncertainties is presented.
This work includes a full treatment of available experimental correlated
systematic errors for both new and old data sets and a systematic and
pragmatic treatment of uncertainties of the parton distributions and their
physical predictions. The new gluon distribution is considerably harder than
that of previous standard fits.  Extensive results on the uncertainties of
parton distributions at various scales, and on parton luminosity functions at
the Tevatron RunII and the LHC, are obtained. The latter provide the means to
quickly estimate the uncertainties of a wide range of physical processes at
these high-energy hadron colliders, such as the production cross sections of
the $W,Z$ at the Tevatron and the LHC, and that of a light Higgs.}

\section{Introduction}

\label{sec:Intro}

Progress on the determination of the parton distribution functions (PDF's) of
the nucleon, from global quantum chromodynamics (QCD) analysis of hard
scattering processes, is central to precision standard model (SM)
phenomenology, as well as to new physics searches, at lepton-hadron and
hadron-hadron colliders. There have been many new developments in recent
years, beyond the conventional analyses that underlie the widely used PDF's
\cite{Mrst0,Cteq}. These developments have been driven by the need to
quantify the uncertainties of the PDF's and their physical predictions$\;$\cite
{Alekhin,Botje,ZomerDis,GKK,Multivariate,Hesse,Lagrange}. This report
describes a comprehensive new global QCD analysis based on the most current
data, and on recently developed techniques of analysis$\;$\cite
{Multivariate,Hesse,Lagrange} that:

\begin{Simlis}
\item  allow the precise treatment of fully correlated experimental
systematic errors in the $\chi ^{2}$ minimization process, thereby greatly
reducing the degree of
difficulty of the problem (to the same as the simple case of no correlated
systematic errors). In the global analysis context, typically involving $%
\sim 2000$ data points and $\sim 100$ different sources of systematic errors,
the conventionally used numerical methods inevitably become practically
untenable and numerically unreliable. (Sec. 2.)

\item  significantly expand the traditional paradigm (in which specific
subjectively chosen ``best fits'' are produced) to a systematic procedure to characterize
the parton parameter space in the neighborhood of the global minimum,
thereby enabling the systematic exploration of the uncertainties of parton
distributions and their physical predictions due to known experimental
errors and uncertainties of input theoretical model parameters. This is made
possible by an iterative procedure to reliably calculate the Hessian matrix
for error propagation in the global analysis context. (Sec. 3)
\end{Simlis}

\noindent Results of this analysis consist of:

\begin{Simlis}
\item  a new generation of parton distributions (CTEQ6), which include the
standard sets CTEQ6M (msbar scheme), CTEQ6D (DIS scheme), and CTEQ6L
(leading order);

\item  uncertainties on parton distributions (embodied in 40 sets of
eigenvector parton distribution sets) and their physical predictions,
e.g.\ uncertainty ranges of various quark-quark, quark-gluon, and
gluon-gluon luminosity functions at the Tevatron and LHC energies---from
which the uncertainties of a variety of standard model and new physics
processes can be inferred.
\end{Simlis}

\noindent Details of this work, as well as extensive references (which do not
fit in this short summary) can be found in Ref.\ [10].

\section{Experimental input, new method of $\protect\chi ^{2}$ minimization, and new CTEQ parton
distributions}

Among the new data sets used in this analysis, the most notable ones are from
 H1$\;$\cite {H1}, ZEUS$\;$\cite{ZEUS}, and D\O$\;$\cite{D0jet}. The greater precision
and expanded $(x,Q)$ ranges compared to previous data in both processes
provide improved constraints on the parton distributions. For the first time
in a full global analysis, the correlated systematic errors for all DIS
experiments are taken into account. Since $\sim 2000$ data points from 15--20
diverse experimental data sets are used in this global analysis, traditional
methods of $\chi ^{2}$ minimization and error propagation are inadequate. In
the covariance matrix approach, numerical instability arises from the
inversion of large dimensional matrices for some data sets. An alternative approach is to add experimental fitting parameters,
one for each source of systematic error, to the theory model (PDF)
parameters. In this case, the total number of fitting
parameters becomes so large (of the order of 100) that general programs of $%
\chi ^{2}$ minimization (such as MINUIT) do not consistently yield reliable
results (with errors).

We overcome this problem$\;$\cite{Hesse,cteq6} by performing the minimization
with respect to the experimental parameters analytically, before the
numerical minimization. This reduces the latter to the same (manageable) level
as the case with only theory parameters. (Cf. Section 2.2 of Ref.\ [10].) A
significant additional advantage in this approach is that the analytic
results on the optimal deviations associated with sources of systematic error
provide important insight on the quality of the fits. It also provides a way
to compare data and theory with the effects of systematic
errors explicitly taken into account. These features are explained in Ref.\ [10] (cf.
Appendix B in particular).

A primary result of the analysis is a \emph{standard set} of parton
distributions (the nominal ``best fit'') in the \mbox{\small {$\overline
{MS}$}} scheme, referred to as CTEQ6M. It provides an excellent global fit to
the data sets used. The overall $\chi ^{2}$ for the CTEQ6M fit is $1954$ for
$1811$ data points. Figure \ref{fig:CtqNewOld} shows an overview of the
comparison between the
new PDF's and the previous generation of CTEQ PDF's, the CTEQ5M1 set, at $%
Q=2 $ GeV.\figCtqNewOld In order to exhibit the behavior of the PDF's clearly
for both large and small $x$ in one single plot, we choose the abscissa to be
scaled according to $x^{1/3}$. Correspondingly, we multiply the ordinate by
the factor $x^{5/3}$, so that the area under each curve is proportional to
the momentum fraction carried by that flavor in the relevant $x$ range. We
see that the most noticeable change occurs in the gluon distribution. It has
become significantly harder than in CTEQ5M1 and all MRST PDF sets at all $Q$
scales. This behavior is mainly dictated by the D\O\  inclusive jet data, which lie
in the range $50 < Q < 500$\,GeV and $0.01 < x < 0.5$. (The higher $\eta$ bins
of this measurement allow a higher $x$ reach than the central jet data
from previous measurements.) The hard gluon distribution becomes amplified at
lower $Q$ scales, due to the nature of QCD evolution and the fact that there
is no direct experimental handle on the gluon at large $x$ and low $Q$. The
enhanced gluon at large $x$ is similar to the CTEQ4HJ and CTEQ5HJ
distributions. More detailed comparisons between the new fits, the data sets
used, and other existing fits can be found in Ref.\ [10].

\section{New method of error propagation and uncertainties of parton
distributions and their physical predictions}

There are formidable complications when standard statistical methods are
applied to global QCD analysis to make error estimates based on quantitative
analysis of the behavior of the $\chi ^{2}$ (or, more generally, likelihood)
function in the PDF parameter space. The basic problem is that a large body
of data from many diverse experiments, which are not necessarily compatible
in a strict statistical sense, is being compared to a theoretical model with
many parameters, which has its own inherent theoretical uncertainties as well
as numerical instabilities associated with multi-dimensional integrations. In
recent papers$\;$\cite{Multivariate,Hesse,Lagrange}, we have proposed and applied
a powerful iterative procedure to reliably calculate the behavior of the $%
\chi ^{2}$ function in the \emph{neighborhood} of the global minimum,
overcoming the long-standing difficulties known to plague standard general
programs (such as MINUIT) when applied to this type of problems.  This method
generates eigenvalues and eigenvectors of the Hessian matrix iteratively, as
it seeks the right (i.e.\ physical) step sizes for finite-difference
calculations in the multi-dimensional PDF parameter space. In the end, the
behavior of the global $\chi ^{2}$ function in the neighborhood of the
minimum is encapsulated in $2N_{p}+1$ sets of orthonormal eigenvector PDF's,
where $N_{p}\sim 20$ is the number of free PDF parameters. Details are given
in Ref.\ [7,9].

From these PDF sets we can calculate the best estimate, and the range of
uncertainty, for the PDF's themselves and for any physical quantity that
depends on them. The uncertainty can be computed from the simple master
formula
$
\Delta X=\frac{1}{2}\left( \sum_{i=1}^{N_{p}}\left[ X(S_{i}^{+})-X(S_{i}^{-})%
\right] ^{2}\right) ^{1/2},  \label{eq:DeltaX}
$
where $X$ is the observable and $X(S_{i}^{\pm })$ are the predictions for $X
$ based on the PDF sets $S_{i}^{\pm }$ from the eigenvector basis. As an
illustration, Fig.\ 2 shows the percentage error bands of the
quark--anti-quark and gluon-gluon luminosity functions at the Tevatron and
the LHC as functions of the parton subprocess CM energy.  The uncertainties
of the quark and gluon distributions themselves, as well as the quark-gluon
luminosity functions, are given in Ref.\ [10].
\figLumi

\section{ Concluding Remarks}

There are many complex issues involved in a comprehensive global parton
distribution analysis. Foremost among these on the experimental side is the
``imperfection'' of real experimental data compared to textbook
behavior. For instance, some experimental measurements appear to be
statistically improbable because $\chi ^{2}\:/\:N$ deviates from $1$
substantially more than the expected $\pm \sqrt{2/N}$; or different precision
experimental measurements of the same physical quantities appear to be
statistically incompatible in all regions of the model parameter space. The
methods of Ref.\ [7-9] cannot resolve these problems
completely---no global analysis method can---but the tools developed in this
formalism have brought much progress to the global analysis endeavor, and have
made possible a deeper look into some of these problems. They allow us to
assess the acceptability and compatibility of the affected data sets in more
practical terms, and to suggest pragmatic ways to deal with the apparent
difficulties. These detailed studies were not possible in the conventional
approach. On the theoretical side, the uncertainties on the perturbative QCD
(PQCD) calculations of the various physical processes included in the global
analysis are not easily quantified in a uniform way. Obviously, much work lies
ahead for continued progress in our effort to pin down the parton structure
of the nucleon, and to test the limits of perturbative QCD.

\end{document}